\documentclass[letterpaper]{JHEP3}

\usepackage{amsfonts}
\usepackage{amssymb}
\usepackage{epsfig}
\usepackage{graphicx}
\usepackage{amsmath}
\usepackage{amsxtra}

\def\circa#1{\,\raise.3ex\hbox{$#1$\kern-.75em\lower1ex\hbox{$\sim$}}\,}
\makeatletter

\def\art{\@ifnextchar[{\eart}{\oart}}
\def\eart[#1]#2#3#4#5#6{{\rm #2}, {\em #3  #4} {\rm (#6) #5} ({\em #1})}
\def\hepart[#1]#2{{\rm #2, \em#1}}
\newcommand{\oart}[5]{{\rm #1}, {\em #2  #3} {\rm (#5) #4}}

\newcounter{alphaequation}[equation]
\def\thealphaequation{\theequation\hbox to
0.6em{\hfil\alph{alphaequation}\hfil}}
\def\eqnsystem#1{
\def\@eqnnum{{\rm (\thealphaequation)}}
\def\@@eqncr{\let\@tempa\relax \ifcase\@eqcnt \def\@tempa{& & &} \or
  \def\@tempa{& &}\or \def\@tempa{&}\fi\@tempa
  \if@eqnsw\@eqnnum\refstepcounter{alphaequation}\fi
\global\@eqnswtrue\global\@eqcnt=0\cr}
\refstepcounter{equation} \let\@currentlabel\theequation \def\@tempb{#1}
\ifx\@tempb\empty\else\label{#1}\fi
\refstepcounter{alphaequation}
\let\@currentlabel\thealphaequation
\global\@eqnswtrue\global\@eqcnt=0 \tabskip\@centering\let\\=\@eqncr
$$\halign to \displaywidth\bgroup \@eqnsel\hskip\@centering
$\displaystyle\tabskip\z@{##}$&\global\@eqcnt\@ne
\hskip2\arraycolsep\hfil${##}$\hfil& \global\@eqcnt\tw@\hskip2\arraycolsep
$\displaystyle\tabskip\z@{##}$\hfil
\tabskip\@centering&\llap{##}\tabskip\z@\cr}
\def\endeqnsystem{\@@eqncr\egroup$$\global\@ignoretrue} \makeatother

\def\be{\begin{equation}}
\def\ee{\end{equation}}
\def\bea{\begin{eqnarray}}
\def\eea{\end{eqnarray}}

\newcommand{\vo}{{\cal V}}

\newcommand{\roughly}[1]{\mathrel{\raise.3ex\hbox{$#1$\kern-0.85em
\lower1ex\hbox{$\sim$}}}}

\def\ba{\begin{eqnarray}}
\def\ea{\end{eqnarray}}
\def\be{\begin{equation}}
\def\ee{\end{equation}}

\def\ssG{{\scriptscriptstyle G}}

\def\NL{{\scriptscriptstyle NL}}

\def\C{\mathcal{C}}

\def\O{\mathcal{O}}

\def\V{\mathcal{V}}

\def\nn{\nonumber}

\def\({\left(}
\def\){\right)}

\def\pref#1{(\ref{#1})}

\title{Non-standard primordial fluctuations\\
and Nongaussianity in String inflation }

\author{
C.P.~Burgess,${}^{1,2}$ M.~Cicoli,${}^3$ M.~G\'omez-Reino,${}^{4,5}$  F.~Quevedo${}^{6,7}$
G.~Tasinato,${}^8$  I.~Zavala${}^9$\\

$^1$ Department of Physics \& Astronomy, McMaster University,
 Hamilton ON, Canada.\\
$^2$   Perimeter Institute for Theoretical Physics,
 Waterloo ON, Canada.\\
$^3$ Deutsches Elektronen-Synchrotron DESY, Hamburg, Germany.\\
$^4$ Theory Division, CERN, CH-1211 Gen\`eve 23, Switzerland.\\
$^5$ Department of Physics, University of Oviedo,
Avda. Calvo Sotelo 18,
Oviedo, Spain.
\\
$^6$ DAMTP/CMS, University of Cambridge, 
 Cambridge CB3 0WA, UK.\\
$^7$ Abdus Salam ICTP, Strada Costiera 11, Trieste 34014, Italy.\\
$^8$ Institut f\"ur Theoretische Physik, Universit\"at
Heidelberg,\\
 \qquad\qquad\qquad\qquad
  Philosophenweg 16 and 19, 69120 Heidelberg, Germany.\\
$^9$ Bethe Center for Theoretical Physics and Physikalisches
Institut\\
 \qquad\qquad\qquad\qquad
 der Universit\"at Bonn, Nu\ss allee 12, D-53115 Bonn,
Germany. }

\date{}

\abstract { Inflationary scenarios in string theory often involve
a large number of light scalar fields, whose presence can enrich
the post-inflationary evolution of primordial fluctuations
generated during the inflationary epoch. We provide a simple
example of such post-inflationary processing within an explicit
string-inflationary construction, using a K\"ahler modulus as the
inflaton within the framework of LARGE Volume Type-IIB string flux
compactifications. We argue that inflationary models within this
broad category often have a selection of scalars that are light
enough to be cosmologically relevant, whose contributions to the
primordial fluctuation spectrum can compete with those generated
in the standard way by the inflaton. These models consequently
often predict nongaussianity at a level, $f_{\NL} \simeq \O(10)$,
potentially observable by the Planck satellite, with a bi-spectrum
maximized by  triangles with squeezed shape in a string
realization of the curvaton scenario. We argue that the
observation of such a signal would robustly prefer string
cosmologies such as these that predict a multi-field dynamics
during the very early universe.
\\

 {\em Dedicated to Lev Kofman, whose untimely passing
 makes the Universe a little bit darker.}}

\preprint{DAMTP-2010-44 \\ DESY 10-071 \\ IC/2010/019 \\ CERN-PH-TH/2010-136 }

\begin{document}

\section{Introduction}

Standard Hot Big Bang cosmology provides a good description of the
great wealth of large-scale observations \cite{CosmoObs} that have
recently revolutionized our understanding of cosmology, but it
only does so if the universe is started off with a particular kind
of initial conditions. Cosmic inflation \cite{Inflation} was
initially proposed as an elegant way of obtaining these conditions
as the outcome of still-earlier dynamics. But this initial promise
was subsequently reinforced by the observation that curvature
perturbations generated by quantum fluctuations of the inflaton
field can get imprinted on the temperature distribution of the
Cosmic Microwave Background (CMB) in the much later universe, in
good agreement with the almost-scale-invariant and Gaussian
spectrum that is observed.

Obtaining the desired inflationary expansion within a realistic
picture of the at-present ill-understood dynamics appropriate to
the very high energies required proved to be much harder than
expected, however. The last decade has seen some progress, sparked
by the understanding of modulus stabilization within string
theory. This allows the construction of calculable inflationary
configurations within string theory, with the role of the inflaton
played either by an open-string degree of freedom --- such as the
relative positions of BPS branes \cite{BraneInf}, or of a brane and
antibrane \cite{BBI}, or Wilson lines \cite{WLs} --- or a field from the closed-string sector --- such as a geometrical modulus \cite{ModInf} (see \cite{SIrevs, MyRev} for reviews).

Nowadays, various string inflationary models are under reasonably
good theoretical control, and developed to a level that can be
compared meaningfully to cosmological data. In particular, because
mechanisms now exist to stabilize moduli, it is possible to
understand the cosmological evolution of {\em all} of the relevant
fields, and therefore to be sure that the motion of fields other
than the inflaton do not ruin the simplest single-field
inflationary predictions for the evolution of curvature
perturbations. It is largely the removal of this potential
theoretical error that now makes the predictions of string
inflationary scenarios sufficiently reliable for comparisons with
observations.

One feature common to the string inflationary models explored so
far is the effective absence in them of isocurvature fluctuations
in the predictions for CMB observables. This despite the fact that
most scenarios involve more than one potentially cosmologically
active scalar field during the inflationary epoch. Indeed models
are usually designed this way, with all of the non-inflaton moduli
sitting in their local minima as the inflaton rolls. Such
constructions greatly simplify the calculation of late-time
perturbations, because they predict only adiabatic fluctuations,
which can be evolved forward to the present time with minimal
sensitivity to the details of the poorly-understood cosmological
history between inflation and now. It is because of this that the
implications of these models are usually well-captured, {\em ex
post facto}, by simple single-field inflationary models
\cite{MyRev, SIsinglefield}.

An unfortunate consequence of the focus for convenience on such
models is the misconception that string inflation {\em must} agree
in its predictions with single-field models, including in
particular a prediction of vanishingly small nongaussianity. This
prediction is sometimes held up as a potential observational way
to discriminate \cite{CyclicNG} between string inflation and
alternatives to inflation within string theory \cite{Cyclic}.

In order to investigate the robustness of such predictions, in
this paper we take the first steps towards exploring other
mechanisms for generating primordial fluctuations within a
concrete string inflationary model based on a LARGE Volume
(LV) scenario. (For other discussions of nongaussianities in string inspired scenarios see \cite{Others}.)
We find we are able to construct such string inflationary
frameworks by making nontrivial use of the presence of the large
number of scalar fields that are generically present during and
after the inflationary epoch. If these fields are sufficiently
light during inflation they can acquire significant isocurvature
fluctuations which post-inflationary evolution can robustly
convert into adiabatic perturbations, swamping those
contributions coming from the inflaton field itself. Although
somewhat more history-dependent than is the standard mechanism,
the subsequent evolution of the resulting adiabatic fluctuations
remains plausibly independent of the details of cosmic evolution
provided only that the universe comes to thermal equilibrium
shortly after adiabatic perturbations with the desired features
are produced.

Our search for models uses two generic mechanisms for achieving
post-inflationary isocurvature to adiabatic conversion: the
curvaton mechanism \cite{LW,Moroi,Enqvist}; and the modulation
mechanism \cite{DGZ1,DGZ2,Kofman,ZaldarriagaNG} scenarios. In
particular, the main models we present can be regarded as explicit
realizations of the curvaton mechanism within a
string-inflationary framework. The idea that such modulus
dependent effects could contribute to curvature perturbations is
not in itself new. What we accomplish in this work is to achieve
it for the first time in a fully calculable string set-up, where
the required properties are subject to a myriad of constraints
imposed by the underlying UV consistency. The precision of this
kind of setup is a necessary preliminary for asking more detailed
questions about reheating and the ultimate transfer of energy from
the inflaton to observable degrees of freedom (d.o.f.). Similar studies for brane-antibrane inflation allowed the identification of cosmic
strings as a potential late-epoch signature \cite{BBCstrings}, as
well as the utility of warping for channeling energy into the
observed low-energy sector \cite{WReheat}. Recent studies of reheating at the end of closed string inflation similarly reveal the need to set severe constraints on the hidden sector dynamics in order to allow an efficient reheating of the visible sector \cite{Reheating}.

Because the observed primordial fluctuations are not directly
generated by the inflaton their properties in general depend
differently on the various underlying parameters, and are in
particular not tied to the slow-roll parameters in the way
familiar from single-field models. This could ultimately allow
string inflationary models for which the string scale is not in
conflict with the demands of particle physics during the present
epoch (such as the supersymmetry breaking scale) \cite{InfvsSUSY},
although we do not yet have an explicit example which does so.

The most interesting such difference is the generic prediction of
a sizeable level of nongaussianity, $f_\NL \simeq \O(10)$; a level
detectable by the Planck satellite. The underlying imprinting of
the adiabatic fluctuations takes place in the post-inflationary
epoch. It is characterized by a non-linear relation between scalar
and curvature fluctuations, that generates nongaussianities of
local form. The corresponding bi-spectrum, consequently,  is
robustly predicted to be dominated by triplets of momenta that
form long, thin triangles: the so-called squeezed
limit.\footnote{For a recent comprehensive review on
nongaussianities see \cite{paul}.}

In the models studied here the size of the
nongaussianity is a consequence of the properties of the geometry
of the extra dimensions in string theory. But if such
nongaussianity should really be observed with these properties,
they will not tell us about microscopic physics in this much
detail. What they most likely would tell us is that the epoch of
fluctuation generation and its aftermath are described by some
sort of multi-field system similar to the ones we describe.

We perform our search for these mechanisms within the LARGE Volume
(LV) scenario of modulus stabilization for Type IIB string vacua
\cite{LV}. These models are convenient for this purpose for
several reasons. First, they predict the existence of a suite of
moduli, whose masses naturally come with a hierarchical
suppression in different ways by powers of the extra-dimensional
volume, $\V = \hbox{Vol}/\ell_s^6$, in string units \cite{cqs, astro,
GenAnalofLVS, ubernat}. In particular, K\"ahler moduli for small
cycles tend to arise with masses of order $M_p/\V$ while those for
large cycles tend to get masses of the order $M_p/\V^{3/2}$ or smaller. Second, the couplings of these moduli to observable fields at late times
can be plausibly estimated provided these fields are assumed to
reside on a brane (or branes) that wrap the cycles whose volumes
are measured by the various moduli \cite{Reheating, astro, LVSatFiniteT}. Finally, inflationary mechanisms are already known using these models, with the inflaton being either a small cycle \cite{kahlerinfl} or a large one 
\cite{fiberinfl} (see also \cite{Berglund:2009uf}).

To construct our models we splice the frameworks developed in
\cite{kahlerinfl} and \cite{fiberinfl}, using a modulus of a small
(blow-up) cycle as the inflaton, keeping the modulus for a larger
cycle as the (curvaton) field that acquires isocurvature
fluctuations. This construction exploits the fact that these
moduli like to be light relative to the inflaton, and so would
plausibly have extra-Hubble fluctuations imprinted on their
profiles. Moreover, after inflation its decay rate to radiation
has the right value to convert isocurvature modes into adiabatic
fluctuations, with the correct amplitude (and a sizeable level of
nongaussianity). The spirit of our construction is to provide an existence proof for mechanisms of this type within a well-developed, modern string set-up, in which  issues associated with moduli dynamics and stabilization can be analysed. Although at first sight the model may seem contrived, it actually uses the minimal amount of ingredients that
are needed in order to exhibit the effects we are interested
in.

The paper is organized as follows. In \S2,
we briefly review the field content and framework of the LV
compactifications. \S3\ then describes the inflationary setup in
these models, which minimally involve 4 moduli: $\V = \V(\tau_1,
\tau_2, \tau_3, \tau_4)$. These are: a curvaton field, $\tau_1$;
the volume modulus, $\tau_2$ together with a blow-up mode,
$\tau_3$, that provides the standard LV stabilization mechanism for
the volume $\V$; and an inflaton $\tau_4$.\footnote{We apologize
for the slightly opaque notation, which is designed to follow
ref.~\cite{fiberinfl} as closely as possible.} These have the
desired hierarchy of masses if the fluxes are adjusted so that the
volumes are stabilized with the hierarchy $\tau_2 > \tau_1 \gg
\tau_4 > \tau_3$. Because of the LV `magic' this can be done using
hierarchies among the input fluxes that are at most $\O(10)$. \S4\
then gives the $\V$-dependence of the couplings of these fields to
observable d.o.f., which we take to be localized on a
brane wrapping either the curvaton cycle or one of the small blow-up cycles. The curvaton mechanism in this framework is explored in \S5, where it is shown that the $\V$-dependence of the masses and couplings can be such as to produce acceptable adiabatic fluctuations. \S6\ then
explores several choices for underlying parameters to get a feel
for the range that is possible for observables. One of the models
presented in this section predicts $f_\NL \simeq 57$. Our
conclusions are briefly summarized in \S7.

\section{The system under consideration}

We start with a discussion of the system whose inflationary
dynamics is of interest. We follow throughout the conventions of
\cite{fiberinfl}.

\subsection{The field content}
\label{FieldContent}

The model requires us to choose a compactification based on a
Calabi-Yau manifold having at least the following 4 K\"ahler
moduli, whose dynamics are of interest:
\begin{itemize}
\item[i)] A fiber modulus, $\tau_1$, playing the role of curvaton
field and wrapped by a stack of $D7$-branes.\footnote{These
$D7$-branes can support either a visible or a hidden sector in a
very model-dependent way.} The low-energy scalar potential first
acquires a dependence on this field through string loop
contributions sourced by the $D7$-branes \cite{GenAnalofLVS,
stringloops}, and for this reason it likes to remain light during
inflation.

\item[ii)] A base modulus, $\tau_2$, that mainly controls the
overall extra-dimensional volume and which is wrapped by a stack
of $D7$-branes needed to generate the string loop potential for
$\tau_1$.${}^3$ This modulus is heavy during inflation, and
remains well-stabilized at its minimum throughout inflation.

\item[iii)] A blow-up mode $\tau_3$, that is an `assisting field'
required to stabilize the volume ${\cal V}$ at its minimum in the
usual LV way (as in \cite{kahlerinfl}). The potential depends on
it through non-perturbative contributions generated by a stack of
$D7$-branes wrapping $\tau_3$ and supporting a hidden sector that
undergoes gaugino condensation.\footnote{$\tau_3$ cannot support a
visible sector due to the tension between non-perturbative effects
and chirality \cite{blumen}.} It is heavy during inflation, and
its VEV is proportional to the logarithm of the volume.

\item[iv)] A second blow-up mode, $\tau_4$, that plays the role of
the inflaton field, as in \cite{kahlerinfl}. Its non-perturbative
potential is again generated by gaugino condensation on the hidden
sector supported by a stack of $D7$-branes wrapping this
cycle.\footnote{The potential for $\tau_4$ cannot be generated by
an instanton since after inflation it would lead $\tau_4$ to a
regime, $\langle\tau_4\rangle < 1$, where we cannot trust the
effective field theory \cite{Reheating}.} During inflation its VEV
is few times the logarithm of the volume.
\end{itemize}
We finally point out that we shall present two explicit scenarios:
\begin{enumerate}
\item Visible sector wrapped around the curvaton cycle $\tau_1$
(and $\tau_2$ since these two cycles intersect each other): this
is the case with the minimal number of 4 K\"{a}hler moduli and,
due to the location of the visible sector on $\tau_1$, it
maximizes the strength of the coupling of the curvaton to visible
d.o.f., so yielding the largest amount of nongaussianities.
However, this is a non-standard realization of the visible sector
supported on a non-rigid 4-cycle which tends to be stabilized
large (giving rise to a tiny gauge coupling).

\item Visible sector wrapped around a blow-up mode $\tau_5$ which
is heavy during inflation: in this case we need to make the system
a bit more involved including a fifth cycle which can be
stabilized small either in the geometric regime (by string loop
effects as in \cite{GenAnalofLVS}) or at the quiver locus (by
$D$-terms as in \cite{quiver}). The advantage is that now we have
a standard realization of the visible sector on a rigid 4-cycle
whose VEV reproduces the correct order of magnitude of the gauge
coupling. However now the geometric separation between $\tau_1$
and $\tau_5$ reduces the strength of the coupling of the curvaton
to visible d.o.f., so yielding a smaller amount of
nongaussianities.
\end{enumerate}

\subsubsection*
{The compactification}

To have these four moduli we consider a Calabi-Yau three-fold with
a K3 fibration structure controlled by two moduli, $\tau_1$ and
$\tau_2$, together with two additional blow-up modes, $\tau_3$ and
$\tau_4$. We assume the Calabi-Yau volume when expressed as a
function of these moduli has the form \cite{ExplicitK3}:
\begin{equation}
 {\cal V}\,=\,\alpha \left( \sqrt{\tau_1} \tau_2-\sum_{i=3}^4
 \gamma_i\, \tau_i^{3/2} \right) \label{initialVolume} \,.
\end{equation}
The K\"ahler potential (including the leading $\alpha'$
corrections) for the effective low-energy 4D supergravity in this
case is (we work throughout in the 4D Einstein frame):
\be \label{K0pot}
 K\,\simeq\, K_0+\delta K_{\alpha'}\,=\,
 -2\,\ln{\left[\vo+\frac{\hat \xi}{2}\right]}\,,
\ee
where the $\alpha'$ corrections are controlled by the quantity
\be
 \hat \xi\,\equiv \,\frac{\xi}{g_s^{3/2}}
 \,=\,-\frac{\zeta(3)\,\chi(M)}{2 \,g_s^{\frac32}
 \,(2 \pi)^3}
\ee
where $\chi(M)$ is the Euler number of the compact manifold. In
applications we take the quantity $\xi$ to lie in the interval
$(0.1\,,\,1)$.

In the superpotential we neglect non-perturbative contributions
associated with the large cycles, $\tau_1$ and $\tau_2$, relative
to those of the small cycles,
\be \label{W0pot}
 W \simeq W_0 + A_3 e^{- a_3 T_3} + A_4 e^{- a_4 T_4} \,,
\ee
since these are negligible relative to those explicitly written
and are likely to be absent since $\tau_1$ and $\tau_2$ are non-rigid cycles. The superpotential is characterized by the constant $W_0$ and the non-perturbative corrections are weighted by constants $A_i$. We
choose the quantity $W_0$ -- as usual in LV models -- to
be order one, and the parameters $a_i$ satisfy
$a_i = 2\pi/N $ since they arise due to gaugino condensation on
$D7$ branes (with $N$ being the rank of the associated gauge group).

Following the LV program, our interest is in the form of the
resulting potential in a regime where $\ln \V \simeq \O\left(
\tau_3 \right)$, so that terms in the $\alpha'$ expansion compete
with the leading non-perturbative contributions from $W$
\cite{cqs}. However, for the inflationary analysis our interest is
not in the local LV minimum. Instead we seek nearby flat
regions of the potential along which the potential is shallow as a
function of $\tau_1$ and $\tau_4$, with $\mathcal{V}$ and $\tau_3$
heavy enough to sit at their local minima. Following the reasoning
of refs.~\cite{kahlerinfl} and \cite{fiberinfl}, we expect such a
regime to arise in the region of field space where the fields are
hierarchically different: $\tau_2 > \tau_1 \gg \tau_4 > \tau_3$,
since in this region the potential likes to become independent of
$\tau_1$ and $\tau_4$, at least before string-loop contributions
are included.

We now use these expressions to compute the scalar potential and
kinetic terms in the desired regime.

\subsection{The kinetic terms}

In this section we investigate the field redefinitions needed to
put the kinetic terms into canonical form. The starting point in
the regime of interest is the K\"ahler metric for the moduli,
which is given by the following symmetric matrix:
\begin{equation}
 K_{i\bar{\jmath}}^{0}=\frac{1}{4\tau _{2}^{2}}\left(
 \begin{array}{cccc}
 \frac{\tau _{2}^{2}}{\tau _{1}^{2}} & \frac{\gamma _{3}
 \tau_{3}^{3/2}+\gamma _{4}\tau _{4}^{3/2}
 }{\tau_{1}^{3/2}} & -\frac{3\gamma _{3}}{2}
 \frac{\sqrt{\tau _{3}}}{\tau _{1}^{3/2}}
 \tau_{2} & -\frac{3\gamma _{4}}{2}
 \frac{\sqrt{\tau _{4}}}{\tau _{1}^{3/2}}
 \tau _{2}  \\
 ^{\prime \prime } & 2 & -3\gamma _{3}
 \frac{\sqrt{\tau _{3}}}{\sqrt{\tau _{1}}
 } & -3\gamma_{4}
 \frac{\sqrt{\tau _{4}}}{\sqrt{\tau _{1}}}   \\
 ^{\prime \prime } & ^{\prime \prime }
 & \frac{3\alpha \gamma _{3}}{2}\frac{
 \tau_{2}^{2}}{\mathcal{V}\sqrt{\tau_{3}}}
 & \frac{9\gamma _{3}\gamma _{4}}{
 2} \, \frac{\sqrt{\tau_{3}\tau_{4}}}{\tau_{1}}  \\
 ^{\prime \prime } & ^{\prime \prime }
 & ^{\prime \prime } & \frac{3\alpha
 \gamma_{4}}{2} \frac{\tau_{2}^{2}}{
 \mathcal{V}\sqrt{\tau_{4}}}
 \end{array}
 \right) ,  \label{LaDiretta}
\end{equation}
where (as in \cite{fiberinfl}) we systematically drop terms that
are suppressed relative to the ones shown by factors
$\sqrt{\tau_i/\tau_2}$ $\forall\,i=3,4$.

The kinetic Lagrangian to leading order therefore
becomes\footnote{We use units with $8 \pi M_p = 1$ unless
otherwise stated.}
\bea - \frac{{\cal L}_{kin}}{\sqrt{-g}}
 &=& \frac{1}{4 \tau_1^2}\left(
 \partial \tau_1\right)^2 +\frac{1}{2
 \tau_2^2}\left( \partial \tau_2\right)^2
 +\sum_{i=3}^4 \frac{3 \alpha \gamma_i}{8
 {\cal V} \sqrt{\tau_i}} \left( \partial
 \tau_i\right)^2+\sum_{i=3}^4 \frac{\gamma_i\tau_i^{3/2}}{2
 \tau_2^2\tau_1^{3/2}} \, \partial
 \tau_1\partial\tau_2  \nonumber\\
 && \qquad - \sum_{i=1}^4 \frac{3 \alpha \gamma_i\,
 \sqrt{\tau_i}}{2 {\cal
 V}\, } \,\left(\frac{\partial \tau_1}{2\tau_1}+
 \frac{\partial \tau_2}{\tau_2}
 \right) \partial \tau_i + \frac{9 \alpha^2 \gamma_3
 \gamma_4}{4}\,\frac{\sqrt{\tau_3
 \tau_4}}{\vo^2}\, \partial \tau_3 \partial \tau_4 \nn\\
 &=&\frac{3}{8
 \tau_1^2}\left( \partial \tau_1\right)^2
 +\frac{1}{2 {\cal V}^2}\left( \partial
 {\cal V}\right)^2 + \sum_{i=3}^4
 \frac{3 \alpha \gamma_i}{8 {\cal
 V} \sqrt{\tau_i}} \left( \partial \tau_i\right)^2
 - \frac{1}{2 \tau_1 {\cal V}}\,\partial \tau_1
 \partial {\cal V} \nonumber \\
 && \qquad - \sum_{i=3}^4 \frac{3 \alpha \gamma_i
 \,\sqrt{\tau_i}}{2 {\cal V}^2\, } \,\partial \vo
 \partial \tau_i + \frac{9 \alpha^2 \gamma_3
 \gamma_4}{4}\,\frac{\sqrt{\tau_3
 \tau_4}}{\vo^2}\, \partial \tau_3 \partial \tau_4\,,
 %
 \label{LagrangKin}
\eea
where the last equality trades $\tau_2$ for ${\cal V}$,
in the limit in which $\tau_1$, $\tau_2$ are much larger
than $\tau_3$, $\tau_4$.
  It is
convenient to canonically normalize order by order in $1/{\cal
V}$, and so we rewrite (\ref{LagrangKin}) as:
\be
 {\cal L}_{kin} = {\cal L}_{kin}^{\mathcal{O}(1)}+{\cal
 L}_{kin}^{\mathcal{O}(\vo^{-1})}+{\cal
 L}_{kin}^{\mathcal{O}(\vo^{-2})} \,, \label{LagKin}
\ee
where the leading term is
\be
 - \frac{{\cal L}_{kin}^{\mathcal{O}(1)}}{\sqrt{-g}}
 =\frac{3}{8 \tau_1^2}\left( \partial \tau_1\right)^2
 +\frac{1}{2 {\cal V}^2}\left( \partial
 {\cal V}\right)^2 - \frac{1}{2 \tau_1 {\cal V}}
 \,\partial \tau_1 \partial {\cal V} \,,
 \label{LagKinO(1)}
\ee
while the subleading terms are
\be
 - \frac{{\cal L}_{kin}^{\mathcal{O}(\vo^{-1})}}{\sqrt{-g}}
 =\sum_{i=3}^4 \frac{3 \alpha \gamma_i}{8 {\cal V}
 \sqrt{\tau_i}} \left( \partial \tau_i\right)^2
 -\sum_{i=3}^4 \frac{3 \alpha \gamma_i\,
 \sqrt{\tau_i}}{2 {\cal V}^2\, } \,\partial \vo
 \partial \tau_i,
 \label{LagKinO(V-1)}
\ee
at $\O(1/\V)$ and
\be
 - \frac{{\cal L}_{kin}^{\mathcal{O}(\vo^{-2})}}{\sqrt{-g}}
 \,=\, \frac{9 \alpha^2 \gamma_3
 \gamma_4}{4}\,\frac{\sqrt{\tau_3
 \tau_4}}{\vo^2}\,
 \partial \tau_3 \partial \tau_4\,,
 %
 \label{LagKinO(V-2)}
\ee
at $\mathcal{O}(\vo^{-2})$. At $\mathcal{O}(1)$ the transformation
\bea
 \tau_1 &=& \exp{\left(a \chi_1 + b \chi_2\right)},
 \label{tau1}\\
 \vo &=& \exp{\left(c \chi_2\right)} \,, \label{vol}
\eea
puts expression (\ref{LagKinO(1)}) into canonical form
\be
 - \frac{{\cal L}_{kin}^{\mathcal{O}(1)}}{\sqrt{-g}}
 =\frac{1}{2}\left[\left( \partial \chi_1\right)^2
 +\left( \partial \chi_2\right)^2 \right] \,,
\ee
where the coefficients $a$, $b$ and $c$ are obtained from the
condition that the matrix $M = \left( \begin{array}{cc} a & b \\ 0
& c \end{array} \right)$ satisfies
\be
    M^{\scriptscriptstyle T} \cdot
 \left( \begin{array}{rrr}
  \frac34 && -\frac12 \\
  -\frac12 && 1 \\
 \end{array} \right) \cdot M
 = I \,.
 \label{condition}
\ee
This has four solutions: $(a,b,c)$, $(a,-b,-c)$, $(-a,b,c)$, $(-a,-b,-c)$, where:
\begin{eqnarray}
 a &=&\frac{2}{\sqrt{3}} ,\text{ \ \ \ \ \ \ \ }
 b = \sqrt{\frac23} ,\text{ \ \ \ \ \ \ \ }
 c =\sqrt{\frac32}\, ,
\end{eqnarray}
and for concreteness we shall choose the first one will all plus signs.
As is shown in \cite{Reheating}, the fields $\chi_1$ and $\chi_2$ turn
out to also diagonalize the mass-squared matrix, $M^2_{ij} =\sum_k
K^{-1}_{ik}V_{kj}$ in the limit where string-loop corrections to
the potential are neglected. Once string-loop corrections are
included a subdominant dependence of $\cal V$ on $\chi_1$ also
arises that is not important for our purposes.

Next we diagonalize the next-order kinetic term,
$\mathcal{L}_{kin}^{\mathcal{O}(\vo^{-1})}$. The first term in
(\ref{LagKinO(V-1)}) becomes diagonal once we rescale the two
small moduli as follows
\be
 \tau_j\,=\,\left(\frac{3 {\cal V}}{4 \alpha \gamma_j}
 \right)^{2/3} \,\phi_j^{4/3}, \text{ \ \
 }\forall\,j=3,4 \label{smalltaus}
\ee
where we use the notation $\phi_j$ with $j=3,4$ to distinguish
these from the large fields, $\chi_1$ and $\chi_2$. The second
term in (\ref{LagKinO(V-1)}) is similarly diagonalized by mixing
$\vo$ with $\tau_j$ $\forall\,j=3,4$. Explicitly, introducing the
following subleading corrections to (\ref{tau1}) and (\ref{vol}):
\bea
 \tau_1 &=& \exp{\left(\frac{2}{\sqrt{3}}
 \, \chi_1 + \sqrt{\frac23} \, \chi_2
 + \frac{3}{2}\sum_{j=3}^4 \phi_i^2\right)}, \label{Tau1}\\
 \vo &=& \exp{\left(\sqrt{\frac32} \, \chi_2
  + \frac{9}{4}\sum_{j=3}^4 \phi_i^2\right)} \,, \label{Vol}
\eea
gives to this order
\be
 {\cal L}_{kin}^{\mathcal{O}(1)}
 +{\cal L}_{kin}^{\mathcal{O}(\vo^{-1})}
 =\frac{1}{2}\sum_{i=1}^2\left(\partial \chi_i\right)^2
 +\frac{1}{2}\sum_{j=3}^4\left(
 \partial \phi_i\right)^2 \,.
\ee
Notice that the last term in eqs.~(\ref{Tau1}) and (\ref{Vol}) is
subleading because $\phi_j \sim \O(\vo^{-1/2}) \ll 1$ for $j=3,4$,
while from (\ref{tau1}) and (\ref{vol}), we have $\chi_i \sim
\O(\ln\vo)$, for $i = 1,2$. We can now substitute (\ref{Vol}) in
(\ref{smalltaus}) to eliminate $\V$ and directly express $\tau_i$
in terms of $\phi_i$, for $i = 3,4$, obtaining
\begin{eqnarray}
 \tau _{i}\, &=&\,\left( \frac{3}{4\alpha
 \gamma_{i}}\right) ^{\frac{2}{3}}
 \left[ \exp {\left(\sqrt{\frac32} \,\chi_{2} +\frac{9}{4}
 \sum_{j=3}^{4}\phi _{j}^{2}\right) }
 \right]^{\frac{2}{3}}\,\phi_{i}^{\frac{ 4}{3}} \nonumber \\
 &\simeq &\left( \frac{3}{4\alpha \gamma_{i}}
 \right)^{\frac{2}{3}}\exp \left[ {\sqrt{\frac23}\,
 \chi_{2} } \right] \, \left( 1+\frac{3}{2}
 \sum_{i\neq j=3}^{4}\phi_{j}^{2}\right)
 \phi_{i}^{\frac{4}{3}},\text{ \ \ }\forall \,i=3,4 \,.
 \label{Smalltaus}
\end{eqnarray}
Notice that passage from the first to the second line neglects
subleading contributions controlled by higher order powers of
$\phi_i$.

Finally, the off-diagonal term in ${\cal
L}_{kin}^{\mathcal{O}(\vo^{-2})}$ is removed by modifying
(\ref{Smalltaus}) slightly, into:
\bea
 \tau _{i}&\simeq& \left( \frac{3}{4\alpha
 \gamma_{i}}\right)^{\frac{2}{3}}\exp \left[
 {\sqrt{\frac23}\,\,\chi_{2} } \right] \,\left(
 1-\frac{3}{4}\sum_{i\neq j=3}^{4}\phi_{j}^{2}\right)
 \phi_{i}^{\frac{4}{3}},\text{ \ \ }\forall \,i=3,4\,,\\
 &\simeq& \left( \frac{3}{4\alpha \gamma_{i}}
 \right)^{\frac{2}{3}} \vo^{\frac23} \,\left(
 1-\frac{9}{4}\sum_{i\neq j=3}^{4}\phi_{j}^{2}\right)
 \phi_{i}^{\frac{4}{3}},\text{ \ \ }\forall \,i=3,4\,.
 \label{SmallTaus}
\eea
The field redefinitions we have determined
render canonical the form of the kinetic terms.

\subsection{The potential}

We next chase these field redefinitions through the definition of
the scalar potential, again following the discussion of
ref.~\cite{fiberinfl}.

\subsubsection*{The potential without loop corrections}

After minimizing the axion directions, the scalar potential
constructed using the K\"ahler potential and superpotential of
eqs.~\pref{K0pot} and \pref{W0pot} (and neglecting subleading
powers of large moduli) is
\be
 V =
\frac{g_s e^{K_{cs}}}{8 \pi}\,
\left[\sum_{i=3}^4\frac{8 \, a_{i}^{2}A_i^2}{3\alpha\gamma_i}
 \left( \frac{\sqrt{\tau_{i}}}{\mathcal{V}}
 \right) e^{-2a_{i}\tau_i}
 -4 \,\sum_{i=3}^4\,W_{0}a_{i} A_i \left( \frac{\tau _{i}}{\mathcal{V}^{2}}
 \right) \, e^{-a_{i}\tau_i}+\frac{3 \,\beta\, \hat\xi W_0^2}{4 \mathcal{V}^{3}}
\right]
 \,.
 \label{ygfdo}
\ee
The overall factor of $g_s e^{K_{cs}}/(8\pi)$ in front of the
potential is a consequence of an overall normalization of the
superpotential, that is needed to express all quantities in the 4D
Einstein frame \cite{LVSatFiniteT}, as is explained in detail in
Appendix \ref{appStringEinstein}.\footnote{From now on we shall
set $e^{K_{cs}}=1$.} The constant $\beta$ appearing in the last,
$\tau$-independent, term,
\be
 V_0\equiv \frac{3 g_s \, \beta \,\hat{\xi}\,
 W_0^2}{32\pi\,\vo^3} \label{defv0}\,,
\ee
includes contributions due to the stabilization of the 
field $\tau_3$, and due to additional uplifting terms needed to uplift the minima 
of the potential from an anti-de
Sitter minimum to a nearly Minkowski vacuum. Its value can be easily found, and
 it is 
 given by the expression
\be
\beta = -\frac{16 \alpha \gamma_4 \langle\tau_4\rangle^{3/2} }{\hat \xi} 
\frac{(1+a_4\langle \tau_4\rangle-2a_4^2\langle\tau_4\rangle^2)}{(1-4 a_4 
\langle \tau_4\rangle)^2}
\ee
where by $\langle \tau_4 \rangle$ we denote the value of the field $\tau_4$ at the minimum.

This potential completely stabilizes $\tau_3$, $\tau_4$ and the
volume ${\cal V}$, at the following values (here we assume $a_i
\tau_i \gg 1$):
\be
 a_i  \langle \tau_i \rangle \,=\,\left(
 \frac{\hat{\xi}}{2 \alpha J} \right)^\frac23,
 \hskip0.7cm  \langle \mathcal{V} \rangle =
 \left( \frac{ 3 \,\alpha \gamma_i }{4a_{i}A_i}
 \right) W_0 \, \sqrt{ \langle\tau_i\rangle }
 \; e^{a_i  \langle \tau_i \rangle },\text{ \ \ \ }\forall\, i=3,4,\label{minhea}
\ee
where $J=\sum_{i=3}^4\gamma_i/a_i^{3/2}$. What is noteworthy is
that eq.~\pref{ygfdo} does not depend at all on the fibre modulus,
$\tau_1$ \cite{fiberinfl}. It does not do so because the dominant
contribution to the potential of large moduli such as these first
arises at the string loop level \cite{GenAnalofLVS}, whose size we
now estimate.

\subsubsection*{The potential with loop corrections}

Each cycle wrapped by a stack of $D7$-branes receives 1-loop open string corrections \cite{GenAnalofLVS, stringloops} which, as pointed out in \cite{fiberinfl}, spoil the flatness of the inflationary potential for $\tau_4$. However it is possible to fine-tune the coefficient of the $\tau_4$-dependent loop correction in order to render it negligible (the amount of fine-tuning needed has been estimated in \cite{Reheating}).
Hence we shall focus only on the $\tau_1$ and $\tau_2$-dependent
loop corrections which can be estimated using a procedure identical to \cite{fiberinfl}:
\bea
 V&=&V_0+\frac{g_s\, a_{4}^{2}A_4^2}{3\pi\,\alpha\gamma_4}
 \left( \frac{\sqrt{\tau _{4}}}{\mathcal{V}}
 \right) e^{-2a_{4}\tau_4}
 -\frac{g_s\,W_{0}\,a_{4} A_4}{2\pi} \left(
 \frac{\tau _{4}}{\mathcal{V}^{2}}
 \right) \, e^{-a_{4}\tau_4}\nonumber \\
 && \qquad\qquad + \left(\frac{A}{\tau_{1}^{2}}
 -\frac{B}{\mathcal{V}\sqrt{\tau_{1}}} +\frac{C\tau_{1}}
 {\mathcal{V}^{2}}\right)\frac{g_s\,W_{0}^{2}}{8\pi\,
 \mathcal{V}^{2}},
 \label{potwl}
\eea
where $A$, $B$, $C$ are given by
\bea
 A&=&\left( g_s C_1^{KK}\right)^2\label{defA}\\
 B&=&4 \alpha C_{12}^{\,W}\label{defB}\\
 C&=& 2\,\left( \alpha g_s\,C_2^{KK}\right)^2 \,,
 \label{defC}
\eea
where $C_1^{KK}$, $C_{12}^{\,W}$, and $C_2^{WW}$ are constants
that depend on the details of the string loop corrections (see
\cite{fiberinfl} for more details). In what follows we regard
these constants as free to be fixed using phenomenological
requirements.

The minimum for $\tau_1$ is at:
\be
  \langle \tau_1 \rangle
 \simeq \left(-\frac{B \vo}{2C} \right)^{2/3}
  \quad \hbox{if $B<0$}
  \qquad \hbox{or} \qquad
 \langle \tau_1 \rangle \simeq
 \left(\frac{4A \vo}{B} \right)^{2/3}
  \quad \hbox{if $B>0$} \,. \label{tau1soln2}
\ee
In the following, for definiteness, we consider the case $B>0$. It
is important to notice that $\langle \tau_1 \rangle$ does not
depend on $\tau_4$, and so $\tau_4$ and $\tau_1$ can evolve
independently in field space. String loop corrections also shift
the minimum for $\tau_3$, with respect to its value in eq.
(\ref{minhea}), but this small correction does not modify the
discussion that follows.

\subsubsection*{The canonically normalized potential}

We next identify that part of the potential relevant to inflation.
We set $\V$ and $\tau_3$ to their minima,  and
follow the dependence of the rest of the potential on the
remaining two fields. This adiabatic approximation is valid in the
large-$\V$ limit because the masses of these fields are
parametrically larger than those of the fields whose motion we
consider.

Recall that the fields $\tau_1$ and $\tau_4$ are given in terms of
their canonically normalized counterparts by
\bea
 \tau_1&=&\vo^{2/3}\,\exp{\left(\frac{2}{\sqrt{3}}\,\langle
 \chi_1 \rangle \right)}\,e^{\frac{2}{\sqrt3}\,\hat\chi_1} \\
 \tau_4 &=& \left( \frac{3 \vo}{4\alpha \gamma_4}
 \,\phi_4^2\right)^{2/3}
 \,\left( 1-\frac{9}{4}\,\phi _{3}^{2}\right) \simeq
 \left( \frac{3 \vo}{4\alpha \gamma_4}\,\phi_4^2\right)^{2/3}
\eea
where we define
\be \label{disc1}
 \chi_1 = \langle \chi_1\rangle
 + \,\hat\chi_1 \,,
\ee
and the approximate equality for $\tau_4$ neglects the subleading
dependence on the modulus $\phi_3$.

Keeping in mind the factors of $g_s$ appearing in the constants
$A$, $B$ and $C$, we expect
\be\label{limtc}
 32 A C\ll B^2\,,
\ee
in weak coupling, and in this case one finds %
\be
 \langle\chi_1
\rangle\,=\,1/\sqrt{3}\,\ln{(q \vo)}, \qquad {\rm with} \qquad
q\equiv 4 A/B\,.
\ee

With this information, the leading contribution to the
inflationary potential breaks into a sum of terms for the would-be
inflaton and curvaton
\be
 V(\phi_4, \hat \chi_1) = V_{inf}(\phi_4)+V_{cur}(\hat\chi_1)
\ee
where
\be \label{inflpot}
 V_{inf}(\phi_4) \simeq V_0-
 \frac{g_s\, W_{0}a_{4} A_4}{2\pi\,\vo^2}
 \left( \frac{3 \vo}{4\alpha \gamma_4}
 \right)^{2/3}\,\phi_4^{4/3} \, \exp{\left\{-\left[
 a_{4} \left( \frac{3 \vo}{4\alpha \gamma_4}\right)^{2/3}\,
 \phi_4^{4/3} \right]\right\}} \,,
\ee
and
\be \label{curvpot}
 V_{cur}(\hat\chi_1) =\frac{g_s\,W_0^2}{8\pi
 \,\vo^{{10}/{3}}}\,\left[
 C_0\,e^{\frac{2}{\sqrt{3}}
 \hat\chi_1}-C_1\,e^{-\frac{1}{\sqrt{3}}\,\hat\chi_1}
 +C_2\,e^{-\frac{4}{\sqrt{3}}\, \hat\chi_1}
 \right] \,,
\ee
with (see  \cite{fiberinfl})
\bea
 C_0&=&C\,q^{2/3}\\C_1&=&B\,q^{-1/3}\\
 C_2&=&A \,q^{-4/3} \,.
\eea
We call $\chi_1$ the curvaton and $\phi_4$ the inflaton because
the potential for $\chi_1$ is parametrically suppressed by powers
of $1/\V$ relative to that for $\phi_4$, thereby ensuring that it
is $\phi_4$ whose energy dominates the cosmic expansion.

Since $\hat\chi_1$ has been defined such that $\hat \chi_1 = 0$ at
the minimum of the potential, it follows that the dependence of
the constants on $\langle \chi_1 \rangle$ ensures, within the
limit (\ref{limtc}), that they satisfy
\be
 \left(\frac{\partial V_{cur}}{\partial
 \hat\chi_1}\right)_{\big{|}
 \,{\hat\chi_1 =0}}\,=\,0\,\,\Rightarrow\,
 C_0+\frac{C_1}{2}-2 C_2\,=\,0\,.
\ee
In the following we work in regimes with $\hat\chi_1$ very small,
for which the exponentials in eq. (\ref{curvpot}) can be expanded
up to quadratic order,
\be\label{curvquad}
 V_{cur}(\hat\chi_1)\, \simeq V_{cur,0} +
 \,\frac{m^2_{\chi_1}}{2}\,\hat\chi_1^2
 \quad {\rm with} \quad
 m^2_{\chi_1}\,=\,\frac{g_s\,W_0^2}{24\pi\, \vo^{10/3}}\,\left[ 4
 C_0-C_1+16 C_2
 \right]\,\equiv\,\frac{g_s\,C_t\,W_0^2}{8\pi\,\vo^{10/3}} \,
\ee
which defines the new constant
\be\label{defct}
 C_t\,\equiv\, \frac{1}{3 }\,\left[ 4 C_0-C_1+16
 C_2 \right] \,.
\ee
The constant piece, $V_{cur,0}$, is absorbable into a subdominant
contribution to the constant $V_0$ in formula (\ref{inflpot}). We
check in our later applications that this quadratic expansion of
the potential  suffices in the regime of interest.

\subsubsection*{Field masses}

For inflationary applications our interest is whether the masses
of the various fields are larger or smaller than the Hubble scale.
Considering that the inflaton potential is of order $V_{inf}
\propto \V^{-2} e^{-a_4 \tau_4} \sim \O(\V^{-3})$, our benchmark
during inflation is $H \sim M_p^2 \,\V^{-3/2}$. Relative to this
consider the following masses, evaluated at the potential's
minimum:

\begin{itemize}
\item If all the fields sit at their minima, the mass spectrum is
(we temporarily reintroduce the dependence on the Planck
mass):
\begin{eqnarray}
 m_{\phi _{i}}^{2} &\sim &\frac{g_s}{4\pi}\,
 \left( \frac{W_{0}}{\mathcal{V}}\right)^{2}
 M_p^{2},\text{ \ \ \ \ \ \ } \forall \,i=3,4 \label{prima}\\
 m_{\chi_2}^{2} &\sim &\frac{g_s W_{0}^{2}}{4\pi
 \,\mathcal{V}^{3}} M_p^{2}, \text{ \ \ \ \ \ \ \ }
 m_{\chi _{1}}^{2}\sim \frac{g_s\,W_{0}^{2}}{
 4\pi\, \mathcal{V}^{3}\sqrt{\tau_{1}}}
 M_p^{2}\sim \frac{g_s\,C_t\,W_0^2}{4\pi\,
 \mathcal{V}^{10/3}}M_p^{2}. \label{seconda}
\end{eqnarray}
\item If the inflation and curvaton fields, $\phi_4$ and $\chi_1$,
are moved away from their minima then their masses are potentially
modified. Inspection of the above formulae shows that the $\chi_1$
mass remains of the same order in $1/\V$ as it is at its minimum,
eq. (\ref{seconda}), while the $\phi_4$ mass changes and becomes
smaller for larger $\phi_4$. Considering, as 
an example, 
 a regime $a_4 \tau_4 \,
> \, \left(2+n\right) \, \ln{\cal V}$, with $n>0$ one finds
\be
 m^2_{\phi_4}\simeq \frac{g_s\,W_0^2}{4\pi\,{\cal
 V}^{3+n}}\,M_p^2\,, \label{minfinf}
\ee
so the inflaton mass is reduced relative to eq. (\ref{prima}) as
its field moves away from its minimum (as in
ref.~\cite{kahlerinfl}).
\end{itemize}

We see from these estimates that for large $\V$, the fields $\chi_2$
and $\phi_3$ have masses that are much larger than $H$, while
$\chi_1$ and $\phi_4$ have masses that are smaller, justifying the
picture wherein $\V$ (which is mostly given by $\chi_2$) and $\tau_3$ (which is mostly $\phi_3$)
can be set to their minima while
both $\chi_1$ and $\phi_4$ remain light enough to have cosmic
fluctuations imprinted on them. Since it is the potential for
$\phi_4$ that is the largest, this is the field whose evolution
dictates the end of inflation and so earns the name inflaton.

\section{Dynamics during inflation}

We next discuss the properties of slow-roll inflation in the above
regime, together with a discussion of whether $\chi_1$ has the
properties required for it to realize the curvaton scenario in
this system. We find these two fields combine the results of
\cite{kahlerinfl} and \cite{fiberinfl}.

We start with the hypothesis that the massive moduli $\chi_2$ and
$\phi_3$ are already at their minima, while $\phi_4$ and $\chi_1$
need not be. We then consider the evolution of the moduli $\chi_1$
and $\phi_4$. As we pointed out before, the analysis is
comparatively simple because these two fields evolve almost
independently: see the potential in eqs. (\ref{inflpot}) and
(\ref{curvpot}). If the field $\phi_4$ acquires a large value, the
dominant term in the inflaton potential is $V_0$. Within this
regime, both $\phi_4$ and $\chi_1$ are lighter than the Hubble
parameter. The former plays the role of inflaton field, while the
latter is the candidate curvaton.

In this section we recap how $\phi_4$ drives inflation, and how
the field $\chi_1$ acquires a scale independent spectrum of
isocurvature fluctuations, of calculable amplitude, during this
inflationary epoch. The next sections discuss how to convert the
isocurvature fluctuations of $\chi_1$ into adiabatic perturbations
after inflation ends.

\subsection{Dynamics of the inflaton field $\phi_4$}

In the scenario just described it is $\phi_4$ that drives
inflation, as in the model of \cite{kahlerinfl}. The inflationary
potential is
\be
 V_{inf}(\phi_4)\,=\,\frac{3 g_s\,\beta\,\hat\xi
 W_0^2}{32 \pi\, \vo^3}-
 \frac{g_s\, W_{0}a_{4} A_4}{2 \pi\,\vo^2}
 \left( \frac{3 \vo}{4\alpha \gamma_4}\right)^{2/3}
 \,\phi_4^{4/3} \, \exp{\left\{-\left[
 a_{4} \left( \frac{3 \vo}{4\alpha \gamma_4}\right)^{2/3}\,
 \phi_4^{4/3}\right]\right\}} \,,
 \label{inflpot2}
\ee
showing again that the scale of inflation is mainly controlled by
the value of the volume, being given by $V_0$ in eq.
(\ref{defv0}).

The corresponding slow-roll parameters, expressed in terms of the
field $\tau_4$, become
\bea
 \epsilon &=& \frac{512 \vo^3}{27\,\gamma_4\,\alpha\,\hat\xi^2
 \beta^2\,W_0^2}\,a_4^2\,A_4^2\,
 \sqrt{\tau_4}\,\left(1-a_4\tau_4 \right)^2
 \,e^{-2 a_4\,\tau_4} \\
 \eta&=&-\frac{16 a_4 A_4\,\vo^2}{9\alpha
\,\hat\xi\,
 \gamma_4\,\beta\,W_0\,\sqrt{\tau_4}}
 \,\left(1-9 a_4 \tau_4 +4 a_4^2 \tau_4^2
 \right)\,e^{-a_4 \tau_4} \,,
\eea
Slow-roll inflation lasts as long as the previous
 slow-roll parameters remain small. In the limit of large
volume, in order to have $\epsilon$ and $\eta$ much smaller 
than one, we choose at horizon exit 

\be  \label{hecons}
 a_4 \tau_4^{in} \gg 2\,\ln{\vo} \,,
\ee
On the other hand, inflation ends at a value $\tau_4^{end}$
such that the $\epsilon$ parameter becomes of order one. 
 These considerations imply that 
   the field range for $\tau_4$ during 
inflation can be parametrized as
\be
 a_4 \tau_4^{end}\,\le a_4 \tau_4\le a_4 \tau_4^{in}
\ee

The number of $e$-foldings is given by the integral
\be\label{nueflds}
 N_e\,=\,\int_{\phi_4^{end}}^{\phi_4^{in}}\,
 \frac{V_{inf}}{V_{inf}'} \; d
 \tilde\phi \,=\,\frac{-9 \alpha \beta\,\hat{\xi}
 W_0\,\gamma_4}{64\,\vo^2\,a_4\,A_4}
 \,\int_{\tau_4^{end}}^{\tau_4^{in}}\,\frac{e^{a_4
 \tau_4}}{\sqrt{\tau_4} \,\left(1-a_4 \tau_4\right)}\,d
 \tau_4\,\ge\,60
\ee

Because we seek the dominant contribution elsewhere, we demand
that the inflaton contribution to the power spectrum of curvature
perturbations is {\em much lower} than the amplitude measured by
the COBE satellite. This gives the following constraint:
\be
 \frac{V_{inf}^{3/2}}{M_p^3\,V_{inf}'}
 \,\ll\,5.2\,\times\,10^{-4} \,.
\ee
Substituting the potential, we find at horizon exit
\be\label{ubcobe}
 \left(\frac{g_s}{8\pi}\right)\, \frac{3^4\,\alpha\,
 \gamma_4\,(\beta\,\hat\xi)^3\,W_0^2}{ 4^6\,a_4^{3/2}
 \,\sqrt{a_4 \tau_4^{in}} \left(1-a_4\tau_4^{in}\right)^2
 } \,\left(\frac{W_0}{ A_4} \right)^2\,
\frac{e^{2 a_4 \tau_4^{in}}}{\vo^{6}}
 \,\ll\,2.7 \times 10^{-7} \,,
\ee

A successful model must satisfy both of the constraints
(\ref{nueflds}) and (\ref{ubcobe}). This imposes conditions on
some of the parameters of the model, which must be supplemented by
the constraints derived in the following sections coming from the
successful realization of the curvaton mechanism. We discuss in
\S6\  explicit scenarios that satisfy all the conditions to
have a successful curvaton model.

\subsection{Dynamics of the curvaton field $\chi_1$}

The curvaton field,\footnote{From now on we drop the {\it hat}
from the field $\hat{\chi}_1$ parameterizing the displacement from
the minimum, in eq. (\ref{disc1}).} $\chi_1$, is lighter than the
Hubble parameter during inflation, since at large volume
\be
 m^2_{\chi_1} \simeq
 \frac{g_s\,C_t\,W_0^2}{8\pi\,\vo^\frac{10}{3}}
 \,\ll\, H_\star^2\,\simeq\,\frac{
  g_s\,\beta\,\hat{\xi}\,W_0^2}{32\pi\,\vo^3} \,,
\ee
where the `$\star$' indicates a quantity evaluated at horizon
exit. During inflation the field $\chi_1$ slowly rolls classically
towards its minimum at zero, but because it is so light it also
undergoes quantum fluctuations that in some circumstances can
dominate the classical motion.

We now estimate when fluctuations dominate, following
\cite{dynamics}. In one Hubble time $H_\star^{-1}$, the light
field $\chi_1$ can fluctuate by an amount $\delta \chi_1 \sim
H_\star/2 \pi$. On the other hand, during the same time interval a
classical slow roll would change the field value by $\Delta \chi_1
\sim - V_{cur}'/(3 H_\star) \, \Delta t_\star = -V_{cur}'/(3
H_\star^2)$. Fluctuations dominate classical evolution\footnote{
This estimate has been debated in the literature, in particular
the value of the power of $H$ in the right hand side of
(\ref{dph}). See for example \cite{Huang}. In this work, we follow
the prescription of \cite{dynamics}, but our approach can be
adapted to different possibilities. We thank Sami Nurmi for
discussions on this point.} when $\delta \chi_1 \sim \Delta
\chi_1$, which occurs when $\chi_1 = \chi_Q$, given by
\be\label{dph}
 V'_{cur}(\chi_Q) \, \simeq\,H_\star^3 \,.
\ee
During inflation quantum fluctuations cause the field $\chi_1$ to
lie in the interval $0 \le \chi_1 \le \chi_Q$ with uniform
probability. Then, its typical value is of order $\chi_1 \sim
\chi_Q$.

In the present case, approximating the curvaton potential as
quadratic, as in eq. (\ref{curvquad}), one finds
\be
 \chi_Q\simeq \left(\frac{g_s}{8\pi}\right)^{1/2}
 \,\frac{(\beta \hat{\xi})^{3/2}\,W_0}{8\,C_t\,\vo^{7/6}} \,,
\ee
and so $\chi_Q \gg \, H_\star$. But because $\chi_Q$ is suppressed
by $1/\vo^{7/6}$ these fluctuations are nevertheless very small at
large volume. A posteriori, it is these powers of $1/\V$ that
justify the expansion of the curvaton potential up to second order
in $\chi_1$.

We now estimate in more detail the amplitude of the power spectrum
for the curvaton fluctuations, following \cite{LUW}. The classical
evolution equation for the curvaton field is
\be
 \ddot{\chi}_1+3 H \dot{\chi}_1+V_{cur}'\,=\,0 \,.
\ee
Making the first order expansion $\delta V_{cur}'\simeq
V_{cur}''\,\delta \chi_1$, one finds the following equation for
the inhomogeneous curvaton fluctuation at superhorizon scales
\be
 \ddot{\delta \chi}_1+3 H\,\dot{\delta \chi}_1
 + V''_{cur}\,\delta \chi_1\,=\,0 \,.
\ee
Since, for a quadratic potential, $\delta \chi_1$ and
$\chi_1$ satisfy the same equation, their ratio does not evolve in
time. This means that this ratio keeps the same value it has at
horizon exit:
\be
 \left(\frac{\delta \chi_1}{\chi_1} \right)
 \,=\,\left(\frac{\delta \chi_1}{\chi_1}
 \right)_\star \,.
\ee
Then the power spectrum of fractional field perturbations
reads
\be
 {\cal P}_{\delta \chi_1/\chi_1}^{1/2}
 \,=\,\frac{H_{\star}}{2\pi\,\chi_\star} \simeq
 \frac{2\,C_t}{\pi\,\beta\,\hat\xi\,\vo^{1/3}}
\ee
where in the last approximate equality we suppose
$\chi_\star\simeq \chi_Q$ (see the previous discussion).

In the next section we discuss how to convert these isocurvature
fluctuations into adiabatic curvature fluctuations when the curvaton decays
after inflation and reheating have already taken place.

\section{Moduli Couplings to Visible Sector Fields}\label{gut}

An important feature of the LV framework is that it is possible to
directly compute the couplings between the moduli (among which the inflaton and the curvaton) and all the other visible or hidden d.o.f.\ localized on $D7$-branes wrapped on internal 4-cycles \cite{Reheating, astro, LVSatFiniteT}. This is a necessary ingredient for calculating the inflaton and curvaton decay rates into visible d.o.f.\ allowing us
to understand reheating at the end of inflation \cite{Reheating}, and to determine whether a curvaton mechanism can be successfully developed.

In the case of the curvaton, we have to focus only on its decay rate to visible gauge bosons. In fact $\chi_1$ is so light that it cannot decay to any supersymmetric particle or even to the Higgs since this receives a large SUSY breaking contribution to its mass. Thus $\chi_1$ can only decay to gauge bosons $g$ and fermions $\psi$ which are massless before the EW phase transition. However it has been shown in \cite{Reheating} that since the fermions are massless, there is no direct decay $\chi_1\to \psi\overline{\psi}$, but only a 3-body decay $\chi_1 \to \psi \overline{\psi} g$ which is suppressed with respect to the 2-body decay
$\chi_1 \to g g$ by a phase space factor. In addition $\chi_1$ cannot decay to light hidden d.o.f.\ since the requirement of a viable reheating forces to have for each hidden sector a pure $N=1$ SYM theory that develops a mass gap \cite{Reheating}.

In order to analyse the coupling of moduli to the gauge bosons of
the field theory living on a stack of $D7$-branes, we proceed as
follows. The $D7$s of interest wrap a 4-cycle whose volume is given
by $\tau$ (which can be any of our moduli): the couplings with the
moduli can be worked out from the moduli dependence of the
tree-level gauge kinetic function $4\pi/g^2 = \tau$ (see
\cite{astro}). In full generality, the kinetic terms read:
\begin{equation}
 \mathcal{L}_{gauge}=-\frac{\tau}{M_p}F_{\mu\nu}F^{\mu\nu} \,,
\end{equation}
and we must expand $\tau$ around its minimum $\tau\to \langle
\tau\rangle +\hat{\tau}$, and go to the canonically normalized
field strength $G_{\mu\nu}$ defined by
\begin{equation}
 G_{\mu\nu}=2\sqrt{\langle\tau\rangle}F_{\mu\nu} \,. \label{redef}
\end{equation}
Doing so, we obtain:
\begin{equation}
 \mathcal{L}_{gauge}=-\frac{1}{4}G_{\mu\nu}
 G^{\mu\nu}-\frac{\hat{\tau}}
 {4 M_p\langle\tau \rangle}
 G_{\mu\nu}G^{\mu\nu} \,. \label{FmunuFmunu}
\end{equation}

\subsection{First scenario}

As explained in section \ref{FieldContent} we imagine the
observable sector to be localized on a stack of $D7$-branes wrapped on the $\tau_1$ and $\tau_2$ cycle, and analyze the decay of the inflaton and curvaton fields into visible gauge bosons. This set-up has the following advantages:
\begin{enumerate}
 \item It represents the simplest example of multi-field curvaton scenario with the smallest number of K\"{a}hler moduli which is 4;
 \item The geometric localization of the visible sector on $\tau_1$
 maximizes the strength of the coupling of the curvaton to visible gauge bosons.
 As we shall see in section \ref{PostInflDynamics}, this will yield the largest amount of nongaussianities.
\end{enumerate}
However there are also some shortcomings:
\begin{enumerate}
\item The K3 fiber is not a rigid cycle and so one has to worry about how to fix the $D7$-brane deformation moduli that would give rise to unwanted matter in the adjoint representation. Here we shall assume that these moduli can be fixed by the use of background fluxes.
\item There is a constraint on the volume of $\tau_1$
coming from constraints on the size that is expected for the
observed gauge coupling. Denoting the gauge coupling as $g$,
using eq. (\ref{tau1soln2}), we have
\begin{equation}
 \frac{4\pi}{g^2}=\tau_1 \simeq
 \,\left(\frac{4 A\,\vo}{B} \right)^{2/3}
 \,.  \label{NEWtau1soln2}
\end{equation}
Focusing for definiteness on a GUT theory,\footnote{Assuming that
the gauge bosons on $\tau_2$ decouple from the EFT getting an
$\mathcal{O}(M_s)$ mass.} $4\pi/g^2\simeq 25$, we find constraints
on the parameters that characterize the string loop contributions.
Indeed, the previous relation implies
\be
 \frac{4 A}{B}\,=\,\frac{125}{\vo}
\ee
from which, using the definitions of $A$ and $B$ in eqs
(\ref{defA}), (\ref{defB}), we obtain
\be
 \left( C_1^{KK}\right)^2\,=\,\frac{125
 \,\alpha}{g_s^2}\,\frac{C_{12}^W}{\vo}
\ee
As we see in the following, when discussing explicit examples,
this condition is relatively easy to satisfy. We do not have to
choose unnaturally large hierarchies between the parameters
$C_1^{KK}$ and $C_{12}^W$.
\end{enumerate}

As studied in \cite{Preheating}, at the end of inflation, due to
the steepness of the potential, the inflaton $\tau_4$ stops
oscillating just after two or three oscillations due to an
extremely efficient non-perturbative particle production of
$\tau_4$ fluctuations. Expanding the canonical normalization
(\ref{SmallTaus}) around the global minimum
($\tau_i=\langle\tau_i\rangle+\hat\tau_i$ $\forall i$) we find
\cite{Reheating}:\footnote{The subleading dependence on
$\hat\chi_1$ is introduced once string loop corrections are
included.}
\be \hat\tau_4\sim \mathcal{O}(\vo^{-1/3})
\hat\chi_1+\mathcal{O}(1) \hat\chi_2+\mathcal{O}(\vo^{-1/2})
\hat\phi_3+\mathcal{O}(\vo^{1/2}) \hat\phi_4,
\ee
realising that the Universe is mostly filled with
$\hat\phi_4$-particles plus some $\hat\chi_2$ and fewer
$\hat\chi_1$ and $\hat\phi_3$-particles. Therefore the energy
density of the Universe is dominated by $\hat\phi_4$ whose decay
to visible d.o.f.\ is responsible for reheating.

The following table summarizes the moduli couplings to visible
gauge bosons living on $\tau_1$ (denoting the corresponding field
strength as $F^{(1)}_{\mu \nu}$):

\begin{figure}[ht]
\begin{center}
\begin{tabular}{c||c|c|c}
  & $\hat\chi_1$ & $\hat\chi_2$ & $\hat\phi_i$,
  $\,\forall\,i=3,4$ \\
  \hline\hline
  \\ & & & \vspace{-0.8cm}\\
  $F^{(1)}_{\mu \nu}  F^{(1)\,\mu \nu} $  & $\frac{2}{\,\sqrt3\,M_p}
$
  & $
\sqrt{\frac{2}{3}}\,\frac{1}{M_p}
$
  & $\frac{3\,\left(\ln{\vo}\right)^{\frac34}}{2\, a_i\,\vo^{1/2}\,M_p}
$ \\
\end{tabular}\\
\end{center}
\begin{center}
\vspace{0.3cm}{{\bf Table {1}:} Couplings between  moduli and
 gauge bosons for a field theory  on the $\tau_1$
 cycle.} \label{table:QFTon1and2}
\end{center}
\end{figure}

Because the light curvaton field mixes through its kinetic terms
with both $\tau_1$ and $\vo$, one might hope to use the
$\chi_1$-dependence of couplings and masses to use the modulation
mechanism \cite{Kofman, DGZ1} to generate the primordial
fluctuations. Although in the present instance the couplings do
not depend on the fluctuations of $\chi_1$, the masses of the
fields do. However, it turns out that in all cases we investigated
the amplitude of modulation-generated fluctuations is too small to
have interesting cosmological consequences. It is for this reason
that we focus on the curvaton mechanism in the following.

We can now derive the total decay rate of a generic
modulus $\varphi$ into gauge bosons $g$:
\be
\Gamma_{\varphi\to  g g}\,=\,\frac{N_g\,\lambda^2\,m_{\varphi}^3}{64 \pi}\, ,
\ee
where $\lambda$ is the coupling listed in Table 1 and $N_g$ is the total number of gauge bosons:
for definiteness we choose $N_g=12$ as in the MSSM. We obtain, for our set of fields,
\begin{eqnarray}
 \Gamma _{\hat \chi_1\rightarrow  g g}
 &=&\frac{1}{4 \pi}\,\frac{m_{\chi_1}^3}{M_p^2}
  \simeq \frac{M_p}{ \vo^5},
 \label{Gamma11} \\
 \Gamma_{\hat \chi_2\rightarrow  g g}
 &=&\frac{1}{8 \pi}\,\frac{m_{\chi_2}^3}{M_p^2}
 \simeq \frac{M_p }{ \vo^{9/2}},
 \label{Gamma110} \\
 \Gamma _{\hat\phi_j\rightarrow g g}
 &=&\frac{27 \left(\ln{\V} \right)^{\frac32}}{64 \pi}\frac{m_{\phi_j}^3}{ \vo M_p^2}
  \simeq \frac{M_p}{ \vo^4}\,,
 \text{ \ \ \ }\forall \,j=3,4. \label{Gamma12}
\end{eqnarray}
where we have emphasized, in the extreme right, the dominant
volume dependence. Notice that the curvaton decay rate is
suppressed with respect to the inflaton decay rate, in the limit
of large volume. This observation plays an important role in the
viability of the mechanism. The reheating temperature in the
approximation of sudden thermalization turns out to be
\cite{Reheating}: \be T_{RH}\simeq \left(\Gamma
_{\hat\phi_4\rightarrow g g} \,M_p\right)^{1/2} \simeq
\frac{M_p}{\vo^2}. \ee

\subsection{Second scenario}

In this section we briefly present a different brane set-up with
the visible sector localized on a small blow-up cycle, showing
that it is possible to build a curvaton scenario with a standard
realization of the visible sector on a rigid del-Pezzo 4-cycle
without any constraint on the overall volume to keep the gauge
coupling from getting too small (given that the VEV of blow-up
moduli does not depend on $\vo$). However, the blow-up mode
supporting the visible sector cannot be either $\tau_3$ or
$\tau_4$ due to the tension between chirality and non-perturbative
effects \cite{blumen}. Hence we need to introduce a fifth modulus
$\tau_5$ with the following three possible brane set-ups
\cite{Reheating}:
\begin{enumerate}
\item Visible sector built with a stack of $D7$-branes wrapped
around $\tau_5$ which is stabilized in the geometric regime (for
example by string loop effects as in \cite{GenAnalofLVS}). In this
case the inflaton $\tau_4$ is not wrapped by the visible sector.
The inflaton and curvaton total decay rates to gauge bosons scale
as: \be
 \Gamma _{\hat \chi_1\rightarrow  g g}
 \simeq \frac{M_p}{ \vo^{17/3}}, \text{ \ \ \ }
 \Gamma _{\hat\phi_4\rightarrow g g}
 \simeq \frac{M_p}{ \vo^4},\text{ \ \ }\Rightarrow\text{ \ \ } T_{RH}\simeq \frac{M_p}{\vo^2}\,.
\ee

\item Visible sector built with a stack of $D7$-branes wrapped around a combination of $\tau_4$ and $\tau_5$ with chiral intersections only on
$\tau_5$ so that the non-perturbative corrections in $\tau_4$ are not destroyed. In this case the
inflaton $\tau_4$ is wrapped by the visible sector. The inflaton and curvaton total decay rates to gauge bosons scale as:
\be
 \Gamma _{\hat \chi_1\rightarrow  g g}
 \simeq \frac{M_p}{ \vo^{17/3}}, \text{ \ \ \ }
 \Gamma _{\hat\phi_4\rightarrow g g}
 \simeq \frac{M_p}{ \vo^2},\text{ \ \ }\Rightarrow\text{ \ \ } T_{RH}\simeq \frac{M_p}{\vo}\,.
\ee

\item Visible sector built via fractional branes at the quiver locus $\tau_5\to 0$
($\tau_5$ can shrink to zero size by $D$-terms as in \cite{quiver}). The inflaton and curvaton total decay rates to gauge bosons scale as:
\be
 \Gamma _{\hat \chi_1\rightarrow  g g}
 \simeq \frac{M_p}{ \vo^{20/3}}, \text{ \ \ \ }
 \Gamma _{\hat\phi_4\rightarrow g g}
 \simeq \frac{M_p}{ \vo^5},\text{ \ \ }\Rightarrow\text{ \ \ } T_{RH}\simeq \frac{M_p}{\vo^{5/2}}\,.
\ee
\end{enumerate}
It is interesting to notice that, due to the geometric separation
between $\tau_1$ and $\tau_5$, the coupling of the curvaton to
visible gauge bosons is weaker than in the first scenario. This
yields a lower level of nongaussianities since, as we shall see in
section \ref{PostInflDynamics}, $f_{\NL}\propto
\Gamma_{\chi_1}^{1/2}$.

\section{Dynamics after inflation: the curvaton mechanism}
\label{PostInflDynamics}

In this section, we summarize the curvaton mechanism for
converting isocurvature fluctuations into adiabatic, curvature
fluctuations in the above inflationary model. Moreover, we
estimate the resulting level of nongaussianity produced in this
process focusing on the first scenario with the visible sector localized on $\tau_1$. However it is easy to re-formulate our analysis for the second scenario.

\subsection{Amplitude of adiabatic fluctuations}\label{ampliadia}

During the inflationary process, both the inflaton and the curvaton masses are much
smaller than the Hubble parameter. The value of the Hubble parameter is essentially
set by the constant piece $V_0$ in the potential (\ref{ygfdo}), that receives contributions
from the uplifting, and from the stabilization of additional fields. This means that
the energy density of the universe is dominated by $V_0$ during inflation, while the
inflaton and curvaton fields provide negligible contributions. Typically, in our
set-up, the masses of these two fields turn out to be not too different during the slow-roll
inflationary period: the inflaton is only few times more massive than the 
 curvaton. In any case,  during  inflation, the two fields 
 are very light and evolve
 independently one from the other.
 Things change drastically towards the end of
inflation:  slow-roll conditions are violated, and the inflaton field $\phi_4$
reaches a region of the potential, where its mass becomes much larger than the
curvaton one.  In this regime, the inflaton field $\phi_4$ dominates the energy
density of the Universe, while at the same time the curvaton energy density remains
subdominant with mass well below the Hubble scale.

 Because of the $\V$ dependence of the decay rates found
above, the small moduli, $\phi_i$, have the largest decay rate --
see eq. (\ref{Gamma12}) -- and so these moduli are also the first
of the moduli to decay. This decay converts the inflaton energy
density into radiation, after which its energy density falls with
the scale factor like $\rho_\gamma \propto a^{-4}$. Since this is
the dominant component of the energy, after this point the Hubble
parameter falls like $a^{-2}$.

Energy tied up in the curvaton field, on the other hand, need not
fall this fast. For instance, once $H$ falls below the curvaton
field's mass this field starts to oscillate coherently around its
minimum, during which its energy density scales like
non-relativistic matter: $\rho_{\chi_1} \sim a^{-3}$. Since this
is much slower than the energy density of radiation the relative
proportion of curvaton energy to radiation energy can grow while
the curvaton oscillates.

This continues until the curvaton field starts to decay, which
happens once the Hubble parameter becomes comparable to the
curvaton decay rate (that, recall, in our set-up is the most
suppressed: see eq.~(\ref{Gamma11})). At this point the curvaton
energy density also converts into radiation, bringing with it any
isocurvature fluctuations that had been stored in the curvaton
field. This converts the curvaton fluctuations into the adiabatic
fluctuations of the radiation energy density.

The total size of the adiabatic fluctuations inherited by such a
conversion depends on the size of the curvaton energy density
relative to the radiation at the point where the curvaton decays.
Denoting this fraction by $\Omega = \rho_{cur}/\rho_\gamma$, then
in a sudden decay approximation, we find:
\be
\Omega \simeq \left[ \frac{1}{6}\,\left( \frac{\chi _{\star }}{M_{p}}\right)
^{2}\,\left( \frac{m_{\chi }}{\Gamma _{\chi _{1}}}\right) ^{\frac{1}{2}}%
\right] ^{\,a}\simeq \,\left[  \frac{\sqrt{2}}{768}\,\frac{%
g_{s}^{1/2}\,(\beta \hat{\xi})^{3}\,W_{0}}{C_{t}^{5/2}\,\vo^{2/3}}\right]
^{a}\, ,
\label{Omega}
\ee
with:
\be
\left\{
\begin{array}{c}
a=1\text{ \ \ \ \ \ \ for radiation dominance \ \ }\Leftrightarrow \text{ \ \ }\Omega \ll 1,
\\
a=4/3\text{ \ \ \ for curvaton dominance \ \ }\Leftrightarrow\text{ \ \ } \Omega \gg 1.
\end{array}
\right.
\ee
The last equality in (\ref{Omega}) substitutes the value of the
various quantities in the present scenario. The resulting
expression for the power spectrum of curvature fluctuations, in
the limit\footnote{It is easy to re-express each quantity in the
curvaton dominance case.} in which $\Omega \ll 1$, is \cite{LUW}:
\be\label{pscf}
 {\cal P}_\zeta^{\frac12}\,=\,\frac23\,\Omega\,{\cal P}_{\delta
 \chi_1/\chi_1}^{\frac12} \,\simeq\,\frac{\sqrt{2}}{576\, \pi} \,
\frac{g_s^{1/2} (\beta \,
 \hat\xi)^2\,W_0}{ C_t^{3/2}\,\vo} \, .
\ee
Demanding this converted amplitude agree with the amplitude
measured by COBE then gives ${\cal P}_\zeta^{\frac12} = 4.8 \times
10^{-5}$, which imposes the constraint
\be
 C_t^{\frac32} \simeq 16\,\frac{g_s^{1/2} (\beta \,
 \hat\xi)^2\,W_0}{ \vo} \,.
 \label{volco}
\ee

\subsection{Nongaussianities}

Following \cite{LUW}, it is not difficult to provide an estimate
for the amount of nongaussianity predicted in this scenario.
We focus on
nongaussianities of local form
\be\label{locAns}
 \zeta\,=\,\zeta_\ssG +\frac35\, f_{\NL}\,\zeta_\ssG^2 \,,
\ee
where $\zeta_\ssG$ is a Gaussian curvature fluctuation.
This Ansatz is particularly well-suited to the present context,
since there is a non-linear relation between scalar fluctuations,
and curvature perturbations produced after inflation ends.
 In
writing eq (\ref{pscf}),
indeed, we implicitly express  the curvature
fluctuation as a first order expansion in the fluctuation of
$\chi_1$. The complete expression,  generalizing the linear
order relation given in eq. (\ref{pscf}),  allows to
exhibit the non-linear connection between scalar and curvature
fluctuations. Indeed, it
reads
\be\label{gendz}
 \zeta\,=\,\frac{\Omega}{3}\,\frac{\delta
 \rho_{\chi_1}}{\rho_{\chi_1}} \,.
\ee
In our case, since we  work with a  quadratic potential, one finds
that
\be\label{gendr} \frac{\delta
 \rho_{\chi_1}}{\rho_{\chi_1}}\,=\,2\,
 \frac{\delta \chi_1}{\chi_1}
 +\frac{(\delta \chi_1)^2}{\chi_1^2}
\ee
Consequently, including this second order
expansion  in the definition of $\zeta$
of eq. (\ref{gendz}),
and comparing with the Ansatz in (\ref{locAns}),
one can read  the following expression for
$f_{\NL}$:
\be\label{genfnl}
 f_{\NL}\,=\,\frac{5}{4\,\Omega}\,\simeq\,
 \,\frac{679 \,C_t^{5/2}\,g_s^{4}\,\vo^{2/3}}{W_0
 \,\beta^3\,\xi^3}\,=\,
  10^5\,\frac{\left( \beta \xi W_0^2
 \right)^{\frac13}}{g_s^{1/6}\,\vo} \,,
\ee
where in the last step we use relation (\ref{volco}). This
expression quantifies the amount of nongaussianity in this
set-up. Notice that  the size of $f_{\NL}$ is
inversely proportional to the conversion factor $\Omega$. This is
expected: if we decrease the efficiency of the conversion process,
by decreasing $\Omega$, we have at the same time to increase the
ratio $\delta \chi_1/\chi_1$ in order to account for the observed
amplitude of fluctuations (see eqs. (\ref{pscf}) and
(\ref{gendz})-(\ref{gendr})). But in this case, the quadratic
contribution in $\delta \chi_1/\chi_1$, in formula eq.
(\ref{gendr}), becomes important in comparison with the linear
term, implying an increase of nongaussianity.

\smallskip

It is also possible to analyse nongaussianity  beyond the
parameter $f_{\NL}$, for example discussing  the parameters
$\tau_{\NL}$ and $g_{\NL}$ that characterize the trispectrum.
Expressions  for these parameters, in curvaton scenarios, have
been provided in the literature: see for example \cite{Wands10}
for a recent review. For our curvaton model, with quadratic
potential, small decay rate $\Omega$
 and in the sudden decay approximation, one finds
\be \tau_{\NL}\,=\,\frac{36}{25} \,f_{\NL}^2\hskip1cm,\hskip1cm
g_{\NL}\,\simeq\,-\frac{10}{3}\,f_{\NL} \ee
with $f_{\NL}$ given in eq. (\ref{genfnl}). The expression for
$\tau_{\NL}$ is the typical one for models where only one species
contributes to the generation of curvature perturbations. The
value of $g_{\NL}$, being proportional to $f_{\NL}$, turns out to
be too low for being detectable by Planck, given the already
stringent bounds on $f_{\NL}$ from WMAP7 \cite{Komatsu10,Smidt:2010sv}. It
would be interesting to extend the model above such as to find
set-ups in which $\tau_{\NL}$ or $g_{\NL}$ turn out to be large,
e.g. as  in the model discussed in  \cite{Byrnes:2009qy}, 
or in which one obtains a sizeable running of nongaussianity, as
analysed in \cite{Byrnes:2009pe}.

\subsection{Constraints from Big-Bang nucleosynthesis}

Besides the requirements of providing the correct amplitude for curvature perturbations, Big-Bang nucleosynthesis (BBN) imposes further constraints on the curvaton model. This since we must ensure that the curvaton field decays by the time BBN takes place, at around $T_{BBN}\sim 1$ MeV.
In order to satisfy this constraint, we impose the following inequality
\be
\Gamma_{\hat \chi_1 \to  g g} > H_{BBN} \sim 10^{-24} \,\text{GeV}
\ee
Using the expression for $\Gamma$ given in (\ref{Gamma11}), we obtain
\be
\Gamma_{\hat\chi_1\to  g  g} \simeq \left(\frac{g_s^{3/2}C_t^{3/2}\, W_0^3}{16(2\pi)^{5/2}}\right)\frac{M_p}{\vo^5}
\ee
Now, recalling that $C_t \simeq  B\,q^{-\frac13}$ within the approximation we are considering, we get an upper bound on the volume,
\be
{\mathcal V} <
7\times 10^{7} \left(g_s^{3/2} B^{2}A^{-1/2} W_0^{3}\right)^{1/5}\, .
\ee
For standard values of the parameters, this imposes a bound on the volume of order $\vo\le 10^8$.

\section{Explicit set-ups}

The previous sections present the conditions that our system must
satisfy in order to furnish a realization of a curvaton scenario.
In this section, we present two representative parameter choices
that satisfy all the constraints, to get a preliminary sense of
how much observable quantities vary.

There is a simple first observation. The results of the previous
sections suggest that once volumes are too large (and so the
inflationary Hubble scale becomes too low) then it becomes
difficult to obtain adequately large primordial fluctuations using
the curvaton mechanism. Indeed, eq. (\ref{pscf}) cannot be
satisfied for volumes that are too large without requiring other
parameters to acquire unnatural values. For typical values of the
parameters a curvaton scenario has a chance for volumes in the
range $10^{3}\,\le\,\vo\,\le\,10^{8}$. Also, eq.~(\ref{genfnl})
shows that very large volumes are usually associated with
nongaussianities of small size. Obtaining a large $f_{\NL}$ is
therefore easiest when choosing relatively small volumes. Because
the underlying expansion is in powers of $\alpha'/\ell_s^2 \propto
1/\V^{1/3}$ we never allow ourselves to consider volumes smaller
than $\V_{\rm min} \simeq 10^3$.


\subsection{First example: small volume, large $f_{\NL}$}

Consider the following representative choice of parameters:

\begin{figure}[ht]
\begin{center}
\begin{tabular}{c|c|c|c|c|c|c|c}
 $\vo$  & $a_4$ & $\xi$  &
  $g_s$  & $W_0$ & $\alpha$ & $A_4$ & $\gamma_4$ \\
  \hline\hline
  \\  $10^3$ & $\frac{1}{10}$ & $\frac{1}{10}$ & $\frac{1}{100}$
& $\frac{1}{10}$ & $6$ & $\frac{1}{10}$ & $20$ \vspace{-0.8cm}\\
    & & & & & & & 
 \\
\end{tabular}\\
\end{center}
\label{table:fc}
\end{figure}

This example is characterized by not-too-large a volume,
$\vo = 10^3$ in Planck units, and by a relatively small string
coupling, $g_s \simeq 10^{-2}$. Also $a_4 =1/{10}$ corresponds to a gauge
group with large rank in the non-perturbative contribution to the
inflaton superpotential.
 Plugging these parameters in eqs.
(\ref{nueflds}) and (\ref{ubcobe}),  and imposing that inflation
starts when the $\epsilon$ parameter is of order $10^{-4}$,
 we find a sufficient number of
e-foldings  ($N_e \simeq 56$). Moreover, there is a small inflaton
contribution to the amplitude of adiabatic fluctuations (${\cal
P}_\zeta^{inf} \, \simeq \, 10^{-2} \, {\cal P}_\zeta^{COBE}$).
Since the volume is relatively small, the scale of inflation is
fairly high in this example. Next, the conditions of having an
acceptable size for the gauge coupling theory, discussed in
section \ref{gut}, imposes the condition $ C_{12}^W\,=\, 10
\left(C_1^{KK}\right)^2$, which in turn implies $C_t \simeq
C_{12}^{W}$. The COBE normalization condition for the curvaton
fluctuations (\ref{volco}) then fixes $C_t\sim 142$. 
 
The most important feature of this model is the high level of
nongaussianity it predicts: using the previous results we find
\be
 f_{\NL} \simeq 57 \,.
\ee
This value can be slightly changed by tuning the choice of
parameters, but the requirement of satisfying all the constraints
does not leave much freedom in this regard. Consequently the order
of magnitude for $f_{\NL}$ is fairly robust in this scenario with
not too large  volume ($\vo \,=\, 10^3$) and high rank gauge group
($a_4\,= 1/{10}$).

\subsection{Second example: larger volume, smaller $f_{\NL}$}

Choosing a different set of parameters shows how the results
change as the volume grows. Consider the following choice

\begin{figure}[ht]
\begin{center}
\begin{tabular}{c|c|c|c|c|c|c|c}
 $\vo$  & $a_4$ & $\xi$  &
  $g_s$ & $W_0$ & $\alpha$ & $A_4$ & $\gamma_4$ \\
  \hline\hline
  \\  $10^6$ & $\frac{1}{8}$ & $1$ & $\frac{1}{100}$
& $10$ & $10$ & $\frac{1}{10}$ & $10$ \vspace{-0.8cm}\\
    & & & & & & & 
 \\
\end{tabular}\\
\end{center}
\label{table:sc}
\end{figure}

\noindent In this example, the volume is larger with respect to
the previous example, while the string coupling and 
$a_4$ are the same. 
Plugging these parameters in eqs. (\ref{nueflds}) and
(\ref{ubcobe}), with the same criteria
of the previous example,  we find a sufficient number of $e$-foldings in
this model ($N_e \simeq 66$) and a small contribution of the
inflaton sector to the COBE amplitude of adiabatic fluctuations
(${\cal P}_\zeta^{inf}\,\simeq\,10^{-3}\,{\cal P}_\zeta^{COBE}$).
After requiring to have an acceptable gauge coupling, as discussed in section \ref{gut}, and imposing COBE normalization condition (\ref{volco}), we find that $C_t\sim 306$. The amount of nongaussianity in this case is small:
\be
 f_{\NL}\simeq 2\,,
\ee
showing that the value of $f_{\NL}$ strongly depends on the choice
of underlying parameters. Different models characterized by
different volumes, although providing the same amplitude for the
spectrum of adiabatic  fluctuations, nevertheless give very
different values for $f_{\NL}$.

\smallskip

In both the previous examples, the ratio between the
 masses of the  inflaton and the curvaton is comparable
during slow-roll inflation: the former is only few
times more massive than the latter. Instead, towards the
end of inflation   when slow-roll conditions
are violated, the inflaton mass becomes much larger
than the curvaton one.
 These appear
to be  general features of our set-ups,
 and depend on parameters that are fixed by the requirements
of having sufficient e-foldings, and the correct amplitude 
for the power spectrum.
 As we   
  discussed at the beginning of section \ref{ampliadia}, these
features are compatible with the requisites for having a succesful
curvaton mechanism.

\section{Conclusions}

In this paper we use LARGE Volume string compactifications to
construct a controlled string-inflation model that does
not use the inflaton to generate primordial fluctuations. Because
the dynamics cannot be captured by a simple single-field slow
roll, it becomes possible to generate observably large
non-gaussianities. These tend to have the local form in the model
examined because they are generated well after inflation ends.

The key ingredients for any such a scenario are twofold. There
must be other fields, besides the inflaton, with masses $m \ll H$
during the inflationary epoch in order to have isocurvature
fluctuations be generated over extra-Hubble distances. The second
ingredient is a mechanism for converting these isocurvature
fluctuations into adiabatic fluctuations.

We find that both ingredients are possible in the LV scenario. The
hierarchy of volume-suppressed modulus masses enjoyed by this
scenario allows some moduli to have masses that are parametrically
suppressed relative to the Hubble scale during inflation, thereby
providing a source of isocurvature fluctuations.

These states also plausibly have a hierarchy of decay rates into
ordinary matter, assuming that ordinary matter is localized on a
brane that wraps one of the cycles whose moduli appear in the
low-energy theory. This allows the isocurvature mode to first
accumulate as an overall fraction of the total energy density, by
oscillating after the inflaton has decayed to radiation. It can
then itself decay at much later times, converting its fluctuations
into adiabatic perturbations. The resulting picture provides a
realization of the curvaton mechanism for string inflationary
models. The fraction of the energy density carried by the curvaton
is suppressed by powers of $1/\V$, naturally leading this fraction
to be a small (and nongaussianities to be comparatively large --
$\O(10)$ -- if the amplitude is the one observed).

Ultimately, the reason such a construction is possible is because
of the potentially large number of fields that can be
cosmologically active during LV inflation. Indeed, should local
nongaussianity be observed, this is probably what it would be
telling us: the dynamics generating primordial fluctuations likely
involves several cosmologically active fields rather than just
one.

Because additional light fields are present these models can be
expected also to manifest other nonstandard mechanisms for
generating fluctuations, such as the modulation mechanism,
although we do not yet have explicit working examples of this
type. A potential benefit of these kinds of models might be the
ability to lower the string scale while still obtaining acceptably
large primordial fluctuations, since this makes it easier to have
a lower supersymmetry-breaking scale, as seems to be preferred by
particle phenomenology in the later universe. As ever, it would be
useful to know how common such models might be in the string
landscape.

It is worth noticing that even though these scenarios require many moduli to work, this is the generic case in string compactifications. The perspective taken in this article is that simplicity arguments using the minimum number of fields are usually good starting points but may not capture the dynamics of the generic case. Furthermore, contrary to most models of string cosmology, we also consider phenomenological constraints, such as the location of the standard-model brane, the value of the present-day gauge coupling, efficient reheating, and so on. We believe this to be crucial because string theory asks to be more than just a model of inflation: string scenarios must therefore address all observable issues and not only a subset of them. Even with these  constraints, we find it encouraging that non-inflaton generation of primordial perturbations appears possible, consistent with having the right amount of inflation required by later cosmology, agreement with current CMB measurements, with potentially observable features like nongaussianity for future experiments. The imminent start of Planck observations makes these questions timely and worth pursuing.

\section*{Acknowledgments}

We wish to thank Tony Riotto for encouraging us to explore
non-standard ways to generate primordial fluctuations, and to Neil
Constable, Sami Nurmi and Andrew Tolley for helpful discussions.
GT would also like to thank Eran Palti for discussions on related
topics a couple of years ago. Various
combinations of us are grateful for the the support of, and the
pleasant environs provided by, the Cambridge Center for
Theoretical Cosmology, Perimeter Institute, the Kavli Institute
for Theoretical Physics in Santa Barbara, McMaster University and
the Abdus Salam International Center for Theoretical Physics. We
also thank Eyjafjallajokull for helping to provide some of us with
unexpected but undivided time. CB's research was supported in part
by funds from the Natural Sciences and Engineering Research
Council (NSERC) of Canada. Research at the Perimeter Institute is
supported in part by the Government of Canada through Industry
Canada, and by the Province of Ontario through the Ministry of
Research and Information (MRI). MGR acknowledges CERN Theory Division for financial support. IZ was partially supported by the SFB-Tansregio TR33  "The Dark Universe" (Deutsche Froschungsgemeinschaft) and the European Union 7th network program "Unification in the  LHC era" (PITN-GA-2009-237920).

\appendix

\section*{Appendix}

\section{String versus Einstein frame}
\label{appStringEinstein}

The correct prefactor of the scalar potential in 4D Einstein frame
has been explicitly shown in \cite{LVSatFiniteT}. Given that the
K\"ahler potential that reproduces the kinetic terms
for the moduli in 4D Einstein frame is known to be (with $S = e^{-\phi} + i C_0$):
\be
\frac{K_E}{M_p^2} = - 2 \ln \vo_E -
\ln( S + \bar{S}) - \ln \left( -i \int \Omega \wedge \bar{\Omega} \right),
\ee
here we shall briefly review just the derivation of the prefactor of the superpotential
starting from the 10D type IIB supergravity action in string frame
(showing only the relevant terms):
\be
S_{10D}^{(s)} \supset \frac{1}{(2 \pi)^7 \alpha'^4} \int d^{10}x \sqrt{-g^{(s)}_{10}}
\left( e^{-2 \phi}\mathcal{R}_{10}^{(s)}  -
  \frac{G_3 \cdot \bar{G}_3}{2 \cdot 3!} \right).
\ee
The action in Einstein frame is obtained via a Weyl rescaling of the metric of the
form $g_{MN}^{(s)} = e^{\phi/2} g_{MN}^{(E)}$:
\be
S_{10D}^{(E)}\supset \frac{2 \pi }{l_s^8} \int d^{10} x \sqrt{-g^{(E)}_{10}}
\left( \mathcal{R}^{(E)}_{10}
- \frac{ G_3 \cdot \bar{G}_3}{12 \, \textrm{Re } S} \right),
\label{10DSEF}
\ee
where $l_s = 2 \pi \sqrt{\alpha'}$.
The dimensional reduction of (\ref{10DSEF}) from 10D to 4D then yields:
\be
S^{(E)}_{4D} \supset \frac{2 \pi}{l_s^8} \left( \int d^{4} x \sqrt{-g_4^{(E)}} \mathcal{R}_4^{(E)} Vol_E
-  \overbrace{\int d^4 x \sqrt{- g_4^{(E)}} \left( \int d^6 x \sqrt{g_6^{(E)}}
\frac{G_3 \cdot \bar{G}_3}{12 \, \textrm{Re } S}
\right)}^{V_{flux}} \right),
\label{4DSEF}
\ee
where $Vol_E = \int d^6 x \sqrt{g_6^{(E)}}\equiv \vo_E l_s^6$.
Comparing the first term in (\ref{4DSEF})
with the Einstein-Hilbert action $S_{EH} = (M_p^2/2) \int d^4 x \sqrt{-g^{(E)}} \mathcal{R}^{(E)}$,
we find:
\be
M_p^2 = \frac{4 \pi \vo_E}{l_s^2} \quad \textrm{ and } \quad
M_s \equiv \frac{1}{l_s}= \frac{M_p}{\sqrt{4 \pi \vo_E}}.
\ee
Writing the superpotential in 4D Einstein frame as:
\be
W_E = \frac{p}{l_s^2} \int G_3 \wedge \Omega,
\ee
the correct prefactor $p$ can be found from requiring that $V_{flux}$ is reproduced by:
\be
V = \int d^4 x \sqrt{-g^{(E)}_{4}} \,e^{K_E/M_p^2} \left[ K^{i \bar{j}}_E
D_i W_E D_{\bar{j}} \overline{W}_E - \frac{3}{M_p^2} W_E \overline{W}_E  \right],
\ee
obtaining $p = M_p^3/(\sqrt{4 \pi})$. Therefore, including the leading order
$\alpha'$ corrections to $K_E$ and non-perturbative corrections to $W_E$,
the $F$-term scalar potential in 4D Einstein frame can be derived from:
\begin{eqnarray}
\frac{K_E}{M_p^2} &=&-2\ln \left[ \vo_E+\frac{\xi}{2}\left(\frac{S+\bar{S}}{2}\right)^{3/2}\right]
-\ln (S+\bar{S})-\ln \left( -i\int \Omega \wedge \bar{\Omega}\right) , \label{K1} \\
W_E &=&\frac{M_p^3}{\sqrt{4\pi }}\left( \frac{1}{l_s^2}\int
G_3\wedge \Omega +\sum_i A_i\,e^{-a_i T_i^{(E)}}\right). \label{W1}
\end{eqnarray}
Stabilising the dilaton $\langle \text{Re} (S) \rangle = g_s^{-1}$ and the complex structure moduli
via background fluxes at tree-level, (\ref{K1}) and (\ref{W1}) reduce to:
\begin{eqnarray}
\frac{K_E}{M_p^2} &=&-2\ln \left( \vo_E+\frac{\xi}{2 g_s^{3/2}}\right) +\ln \left(\frac{g_s}{2}\right)+K_{cs} , \label{K2} \\
W_E &=&\frac{M_p^3}{\sqrt{4\pi }}\left( W_0 +\sum_i A_i\,e^{-a_i T_i^{(E)}}\right), \label{W2}
\end{eqnarray}
where:
\be
K_{cs}=- \ln \left( -i\int \langle\Omega \wedge \bar{\Omega}\rangle\right),\text{ \ \ \ and \ \ \ }
W_0=\frac{1}{l_s^2}\int \langle G_3\wedge \Omega\rangle.
\ee
Hence the prefactor of the scalar potential in 4D Einstein frame can be worked out from:
\be
e^{K_E/M_p^2}\frac{|W_E|^2}{M_p^2}\text{ \ \ \ }
\Longrightarrow\text{ \ \ \ }\left(\frac{g_s e^{K_{cs}}}{8\pi}\right)M_p^4.
\ee
The expressions for $K_s$ and $W_s$ in 4D string frame can be derived by transforming
the scalar potential (recalling that $T^{(E)}_i=T^{(s)}_i / g_s$), and then working out the form of
$K_s$ and $W_s$ that reproduce such a potential. We obtain:
\begin{eqnarray}
\frac{K_s}{M_p^2} &=&-2\ln \left( \vo_s+\frac{\xi}{2}\right) +\ln \left(\frac{g_s}{2}\right)+K_{cs} , \label{K3} \\
W_s &=&\frac{g_s^{3/2} M_p^3}{\sqrt{4\pi }}\left( W_0 +\sum_i A_i\,e^{-a_i T_i^{(s)}/g_s}\right). \label{W3}
\end{eqnarray}
Thus the prefactor of the scalar potential in 4D string frame can be worked out from:
\be
e^{K_s/M_p^2}\frac{|W_s|^2}{M_p^2}\text{ \ \ \ }
\Longrightarrow\text{ \ \ \ }\left(\frac{g_s^4 e^{K_{cs}}}{8\pi}\right)M_p^4.
\ee

\end{document}